\documentclass[aps,eqsecnum,showpacs,amsmath,amssymb]{revtex4}
\usepackage{graphics}
\begin{document}

\preprint{{DOE/ER/40762-305}\cr{UMPP\#04-031}}

\begin{flushright}
DOE/ER/40762-305\\
UMPP\#04-031
\end{flushright}

\count255=\time\divide\count255 by 60
\xdef\hourmin{\number\count255}
  \multiply\count255 by-60\advance\count255 by\time
 \xdef\hourmin{\hourmin:\ifnum\count255<10 0\fi\the\count255}

\newcommand{\xbf}[1]{\mbox{\boldmath $ #1 $}}

\newcommand{\sixj}[6]{\mbox{$\left\{ \begin{array}{ccc} {#1} & {#2} &
{#3} \\ {#4} & {#5} & {#6} \end{array} \right\}$}}

\newcommand{\threej}[6]{\mbox{$\left( \begin{array}{ccc} {#1} & {#2} &
{#3} \\ {#4} & {#5} & {#6} \end{array} \right)$}}

\title{Pion-Nucleon Scattering Relations at Next-to-Leading Order in
$1/N_c$}

\author{Thomas D. Cohen}
\email{cohen@physics.umd.edu}

\author{Daniel C. Dakin}
\email{dcdakin@physics.umd.edu}

\author{Abhinav Nellore}
\email{nellore@physics.umd.edu}

\affiliation{Department of Physics, University of Maryland,
College Park, MD 20742-4111}

\author{Richard F. Lebed}
\email{Richard.Lebed@asu.edu}

\affiliation{Department of Physics and Astronomy, Arizona State
University, Tempe, AZ 85287-1504}

%\date{\hourmin, \today}
\date{March, 2004}

\begin{abstract}
We obtain relations between partial-wave amplitudes for $\pi N
\rightarrow \pi N$ and $\pi N \rightarrow \pi \Delta$ directly
from large $N_c$ QCD\@.  While linear relations among certain
amplitudes holding at leading order (LO) in $1/N_c$ were derived
in the context of chiral soliton models two decades ago, the
present work employs a fully model-independent framework based on
consistency with the large $N_c$ expansion.  At LO in $1/N_c$ we
reproduce the soliton model results; however, this method allows
for systematic corrections.  At next-to-leading order (NLO), most
relations require additional unknown functions beyond those
appearing at LO and thus have little additional predictive
power.  However, three NLO relations for the $\pi N \rightarrow
\pi \Delta$ reaction are independent of unknown functions and
make predictions accurate at this order.  The amplitudes relevant
to two of these relations were previously extracted from
experiment.  These relations describe experiment dramatically
better than their LO counterparts.

\end{abstract}

%11.15.Pg   Expansion for large numbers of components (e.g., 1/Nc expansion)
%13.75.Gx   Pion-baryon interactions

\maketitle

\section{Introduction \label{sec:intro}}

It is now thirty years since 't~Hooft noted that treating the
number $N_c$ of QCD colors as an expansion parameter yields a
limiting theory with substantial predictive power~\cite{TH}; in
1979, Witten extended this general idea to the generic properties
of baryons~\cite{EW}. During the subsequent years, two main
approaches to understanding the spin and flavor dependence of
baryon properties arose.  The first is the chiral soliton
approach of Witten, Adkins, and Nappi~\cite{EW,ANW,AN} which
reinterprets Skyrme's original soliton idea~\cite{S} in the
context of large $N_c$ QCD.  It was noted early on that many
relations among observables in such models depend only on the
overall structure of the soliton models and are completely
independent of the dynamical details~\cite{AN}.  This suggested
that these relations directly reflect general results of large
$N_c$ QCD. An alternative fully model-independent approach based
on consistent power counting of $N_c$ factors in baryon-meson
scattering processes was invented by Gervais and Sakita~\cite{GS}
and Dashen and Manohar~\cite{DM}, and then systematically
developed by Dashen, Jenkins, and Manohar~\cite{DJM}.  In this
approach, an underlying contracted SU($2 N_f$) spin-flavor
symmetry ($N_f$ being the number of light quark flavors) emerges
as $N_c \rightarrow \infty$.  The apparently model-independent
relations of soliton models then automatically emerge at leading
order in $1/N_c$ as results of the group structure of this
emergent symmetry.  The approach based on large $N_c$ consistency
conditions has two obvious advantages over the soliton approach:
It is manifestly model independent, and it allows for systematic
$1/N_c$ corrections.

The systematic treatment of $1/N_c$ corrections comes at a cost.
As in any effective theory, one must generally add new unknown
coefficients at subleading orders.  The power counting in the
$1/N_c$ expansion implicitly constrains the typical size of these
coefficients via naturalness criteria, but when an unknown
coefficient enters a relation at next-to-leading order (NLO), one
essentially has no predictive power beyond what is seen at
leading order (LO).  However, there exist certain relations that
hold even after the inclusion of NLO coefficients; we denote such
relations as ``gold-plated''.  The gold-plated relations hold at
NLO and, hence, should have errors of $O(1/N_c^2)$ relative to the
$O(N_c^0)$ amplitudes.  Since $1/N_c^2 = 1/9$ for the physical
world, these gold-plated relations may be taken to be
semi-quantitative predictions.  In fact, these relations are
often satisfied quite well.  For example, one predicts $g_{\pi N
\Delta}/g_{\pi N N}= \frac 3 2 \,[1 + O(1/N_c^2)]$~\cite{DJM},
and the experimental value of the ratio deviates from 3/2 by only
a few percent.  In contrast, ordinary ``silver-plated'' relations
(those holding only at LO in $1/N_c$) are typically of a more
qualitative nature.

In this paper we use the large $N_c$ consistency condition
approach to deduce relations among partial-wave amplitudes for
the processes $\pi N \rightarrow \pi N$ and $\pi N \rightarrow
\pi \Delta$, which involve only the two light quark flavors $u$
and $d$.  LO relations were derived long ago in the context of
chiral soliton models~\cite{HEHW,MP}. Since these results were
found to be independent of the dynamical details of any particular
soliton model, it was generally assumed that they are fully
model-independent consequences of large $N_c$ QCD, holding at LO
($N_c^0$) for meson-baryon scattering amplitudes.  It was
recently noted that these LO relations can be obtained directly
from the group structure arising from large $N_c$ consistency
conditions~\cite{CL}.  As observed above, one clear advantage of
this approach is that it provides a straightforward formalism for
working to higher order in $1/N_c$.  If one insists upon
model-independent constraints, one in fact gains no predictive
power at higher order except through gold-plated relations, for
which the LO correction terms cancel.  In this paper, we show
that such gold-plated relations do exist, but they necessarily
involve the process $\pi N \rightarrow \pi \Delta$.  We also show
that these gold-plated relations hold moderately well
experimentally, while the analogous silver-plated relations work
quite poorly.  Thus we find that one can at least
semi-quantitatively understand some aspects of the $\pi N
\rightarrow \pi \Delta$ reaction from {\it ab initio\/} large
$N_c$ QCD considerations.

Before proceeding, it is useful to discuss the original
derivation of the relations among partial-wave amplitudes from
the soliton model and to explain why this result is considered to
hold at LO in $1/N_c$~\cite{MP}.  While we derive in this paper a
general large $N_c$ rather than merely a soliton model result, we
believe that first seeing the result in a concrete realization is
instructive, and makes a connection with the older literature. In
chiral soliton models baryons are supposed to arise from {\it
hedgehog\/} configurations, which are the static, finite-energy
solutions ({\it solitons}) of a (nearly) chirally symmetric
tree-level pion Lagrangian.  Such configurations can be assigned
baryon number unity, but break both isospin and rotational
symmetry while preserving ``grand spin'' $\vec{K} \equiv
\vec{I}+\vec{J}$.  Note that classical configurations are
justified by large $N_c$ considerations, since quantum
fluctuations [associated with the excitation of only a few of the
$O(N_c)$ constituents] contribute only at relative order
$N_c^{-1}$.  Static rotations of such classical soliton
configurations lead to energetically degenerate solutions of the
classical equations, implying the existence of a multiplet of
degenerate states at large $N_c$.  Slow rotational motion (with
angular velocity $\sim N_c^{-1}$) among such states is orthogonal
to intrinsic quantum excitations of the soliton and may be
quantized separately.  This quantization leads to nearly
degenerate physical states with the usual physical quantum
numbers and mass splittings of $O(1/N_c)$.  The underlying
hedgehog structure implies that each physical state has $I = J$,
whose common value we label by $R$.  The $R=\frac 1 2$ states are
identified as nucleons and $R=\frac 3 2$ states are identified as
$\Delta$ resonances for the $N_c=3$ world, while states with
higher values of $R$ are generally assumed to be large $N_c$
artifacts.

Physical pions are treated as fluctuations about the soliton; the
action is expanded perturbatively in the number of pion fields,
which is justified for large $N_c$ since each additional pion
field suppresses the amplitude by a factor $\sim 1/\sqrt{N_c}$.
The scattering is then described in terms of the Green function
of the pion-soliton system.  The standard machinery of
semiclassical projection then allows one to obtain amplitudes for
states with well defined $I=J$ in both the initial and final
states.  The $S$ matrix for such a channel in this formalism is
given by
\begin{eqnarray}
S_{LL^\prime R R^\prime I_s J_s} & = &\sum_K (-1)^{R^\prime - R}
\sqrt{[R][R^\prime]}\,[K] \left\{ \begin{array}{ccc} K & I_s & J_s\\
R & L & 1 \end{array} \right\} \left\{
\begin{array}{ccc} K & I_s & J_s \\ R^\prime & L^\prime & 1
\end{array} \right\}
s_{KLL^\prime} + O(N_c^{-1}), \label{MPeqn} \\
\nonumber
\end{eqnarray}
where $[x] \equiv 2x+1$, $R$ ($R'$) is the spin/isospin of the initial
(final) baryon, $L$ ($L'$) is the relative orbital angular momentum of
the initial (final) pion about the baryon, and $J_s$, $I_s$ indicate
the total spin and isospin, respectively, as measured in the
pion-baryon $s$-channel.  Note that the $S$ matrix is a reduced matrix
element (in the sense of the Wigner-Eckart theorem) in terms of both
angular momentum and isospin, in that dependence on the quantum
numbers $(I_s)_3$ and $(J_s)_3$ has been factored out.

The explicit ``1'' in the $6j$ coefficients arises from the isospin of
the pion.  Although this formula holds for pion scattering, it has
been generalized to mesons with spin one ({\it e.g.}, $\rho$) and/or
isospin zero ({\it e.g.}, $\eta$)~\cite{Mat3}.  In such cases,
Eq.~(\ref{MPeqn}) maintains the same basic form except the $6j$
coefficients are either replaced by $9j$ coefficients to account for
the extra vector (spin 1) or collapse to Kronecker deltas (isospin 0).

The preceding derivation exploits the large $N_c$ limit in
multiple ways.  As noted above, the use of the classical hedgehog
itself is justified only for large $N_c$, so that quantum
fluctuations are relatively unimportant.  Moreover, baryon recoil
is neglected in the scattering process since the baryon mass
scales as $N_c$, while the characteristic scattering energy scale
is $O(N_c^0)$.  Similarly, the rotation of the soliton during the
scattering event is neglected since the soliton moment of
inertia, and hence the rotational period, also scales as $N_c$.
These approximations are only valid to LO in $1/N_c$, and thus
any predictions based on this formalism can be expected to hold
only at LO in $1/N_c$.

The energy-dependent function $s_{KLL'}$ in Eq.~(\ref{MPeqn}) is
called a {\it reduced amplitude\/} and contains all the dynamical
information from the chiral soliton model.  Note that these
reduced amplitudes depend only on three variables, while the
physical amplitudes depend on six: The same underlying soliton
structure contributes to multiple physical states.  It is
precisely because there are fewer reduced amplitudes than
physical amplitudes that one can obtain relations between the
physical amplitudes.  The physical interpretation of the label
$K$ is clear: It labels the grand spin of the given excitation.

Equation~(\ref{MPeqn}) has been used with considerable success to
describe baryon spectroscopy.  One approach uses a particular
soliton model to evaluate explicitly the reduced matrix elements
and then to predict fully the physical scattering amplitudes.  The
detailed behavior of these amplitudes can then be used to predict
values for baryon resonance observables~\cite{WE,MK}.  Recently it
was noted that Eq.~(\ref{MPeqn}) has model-independent
applications in the study of baryon resonances~\cite{CL}.  There,
it was noted that a resonance in a given channel corresponds to a
pole in the $S$ matrix, meaning that this pole must appear in one
of the associated reduced amplitudes.  Since the same reduced
amplitudes occur in multiple scattering channels, degeneracies
must exist in the excited baryon spectrum at leading order
($N_c^0$) in $1/N_c$.

While the prediction of degeneracies in the excited baryon
spectrum at large $N_c$ depends upon there being more $S$ matrix
elements than reduced amplitudes, the same fact implies the
existence of linear relations among scattering
amplitudes~\cite{MP,HEHW}.  This result is made explicit by
algebraically eliminating the reduced amplitudes, yielding linear
relations among the physically measurable amplitudes.  Such
silver-plated relations were derived by Mattis and Peskin
(MP)~\cite{MP} for $\pi N \rightarrow \pi N$ and $\pi N
\rightarrow \pi \Delta$ and are a focus of this paper.  The $\pi
N \rightarrow \pi N$ relations were first noted in the context of
Skyrme models (but not large $N_c$ {\it per se}) by
Ref.~\cite{HEHW}. We present them now for future reference, using
the more compact notation (and noting that real initial target
baryons are always nucleons, $R=1/2$) $S_{L L R R^\prime I_s J_s}
\to S^{\pi R^\prime}_{L, 2I_s, 2(J_s-L)}$, or $S_{L L^\prime R
R^\prime I_s J_s}\to S^{\pi R^\prime}_{L, L^\prime, 2I_s,
2(J_s-L)}$ if $L \neq L^\prime$:
\begin{eqnarray}
S^{\pi N}_{L, 3,-1} & = & \frac{L-1}{4L+2}S^{\pi N}_{L, 1,-1}+
\frac{3(L+1)}{4L+2}S^{\pi N}_{L, 1,+1} + O(N_c^{-1}) , \label{LR1}\\
S^{\pi N}_{L, 3,+1} & = & \frac{3L}{4L+2}S^{\pi N}_{L, 1,-1}+
\frac{L+2}{4L+2}S^{\pi N}_{L, 1,+1} + O(N_c^{-1}) , \label{LR2}\\
S^{\pi \Delta}_{L, 3,-1} & = & \frac{4(L-1)}{\sqrt{10}
(2L+1)}S^{\pi \Delta}_{L, 1,-1}+ \frac{3}{2L+1}
\left[\frac{(L+1)(2L+3)(2L-1)}{10L}\right]^{1/2}S^{\pi
\Delta}_{L, 1,+1} + O(N_c^{-1}) , \label{LR3}\\
S^{\pi\Delta}_{L, 3,+1} & = & \frac{3}{2L+1}\left[\frac{L(2L+3)(2L-1)}{10(L+1)}
\right]^{1/2}S^{\pi\Delta}_{L, 1,-1}+\frac{4(L+2)}{\sqrt{10}(2L+1)}S^{\pi
\Delta}_{L, 1,+1} + O(N_c^{-1}) , \label{LR4}
\end{eqnarray}
which are MP Eqs.~(3.22$a$,$b$) and (3.23$a$,$b$), respectively,
\begin{subequations}
\begin{eqnarray}
\sqrt{L+1} \, S^{\pi \Delta}_{L,L+2,1,+1}& = & - \sqrt{L+2} \,
S^{\pi \Delta}_{L+2,L,1,-1} + O(N_c^{-1}) ,\label{LR5a}\\
\sqrt{L+1} \, S^{\pi \Delta}_{L,L+2,3,+1} & = & - \sqrt{L+2} \,
S^{\pi \Delta}_{L+2,L,3,-1} + O(N_c^{-1}) , \label{LR5b}
\end{eqnarray}
\end{subequations}
\begin{subequations}
\begin{eqnarray}
\sqrt{10(L+1)} \, S^{\pi \Delta}_{L,L+2,3,+1} & = & +
\sqrt{L+2} \, S^{\pi \Delta}_{L+2,L,1,-1} + O(N_c^{-1})
\label{LR5primea}\\
S^{\pi \Delta}_{L+2,L,1,-1} & = & - \sqrt{10} \, S^{\pi
\Delta}_{L+2,L,3,-1} + O(N_c^{-1}) , \label{LR5primeb}
\end{eqnarray}
\end{subequations}
which are MP Eqs.~(3.24) (and only three of the four preceding
relations are independent), and
\begin{equation}
S^{\pi N}_{L, 1,-1}-S^{\pi N}_{L, 1,+1}=\sqrt{\frac{2L-1}{L+1}} \,
S^{\pi \Delta}_{L, 1,-1}+\sqrt{\frac{2L+3}{L}} \,
S^{\pi \Delta}_{L, 1,+1} + O(N_c^{-1}). \label{LR6}
\end{equation}
which is MP Eq.~(3.25).

Since these relations were derived from an underlying expression
that is only justified at LO in $1/N_c$, the relations are {\it a
priori\/} only known to be justified at LO in $1/N_c$.  Our
purpose in this work is to include the effects of NLO corrections
in the $1/N_c$ expansion in order to obtain relations that hold
at this higher order.  As noted above, treatments based on the
soliton model do not easily lend themselves to systematic
higher-order corrections. Accordingly, we work with the formalism
based on large $N_c$ consistency rules.  As a first step, in
Sec.~\ref{sec:derive} we rederive Eq.~(\ref{MPeqn}) [the LO
expression from which Eqs.~(\ref{LR1})--(\ref{LR6}) are obtained]
directly from the model-independent formalism based on large
$N_c$ consistency conditions.  Next, we use the large $N_c$
framework to compute the NLO corrections.  We present three new
relations in Sec.~\ref{lin} and compare them with the
experimental data in Sec.~\ref{exp}.

\section{Derivation \label{sec:derive}}

The relations among $S$ matrix elements for different partial-waves
presented in Sec.~\ref{sec:intro} were derived in the context of a
chiral soliton model in the large $N_c$ limit.  However, it is
possible to show that these relations follow directly from large $N_c$
QCD in a fully {\it model-independent} manner~\cite{CL}.  Of course,
this model independence is not surprising since the relations in
Eqs.~(\ref{LR1})--(\ref{LR6}), although derived in the context of a
soliton model, are completely insensitive to the dynamical details of
the model.  We explicitly demonstrate model independence using the
methods of Ref.~\cite{DJM}, which is useful since the procedure
illuminates the method by which we extend the linear relations to
higher order in $1/N_c$.

The key point is the connection between the $s$-channel
amplitudes of physical interest and the same reactions expressed
in terms of $t$-channel amplitudes.  As discussed below, large
$N_c$ QCD severely limits the form of these $t$-channel
amplitudes~\cite{KapMan}.  Thus, we rewrite Eq.~(\ref{MPeqn}) in
terms of $t$-channel amplitudes rather than $s$-channel exchanges
using the following $6j$ symbol identity~\cite{edmonds}:
\begin{eqnarray}
\! \! \! \! \left\{  \begin{array}{ccc} K & I_s & J_s
\\ R & L & 1 \end{array} \right\} \! \left\{
\begin{array}{ccc} K & I_s & J_s \\ R^\prime & L^\prime & 1
\end{array}\right \} & \! \! \! = \! \! \! &\sum_{\cal J} \,
(-1)^{I_s + J_s + L + L^{\prime} + R + R^{\prime} + K + {\cal J}} \,
[{\cal J}] \left\{
\begin{array}{ccc} 1 & R^\prime & I_s
\\ R & 1 & {\cal J} \end{array} \right\} \!
\left\{ \begin{array}{ccc} L^\prime & R^\prime & J_s \\ R & L & {\cal
J} \end{array} \right\} \! \left\{
\begin{array}{ccc} 1 & L^\prime & K
\\ L & 1 & {\cal J} \end{array} \right\} \! .
\label{ident}
\end{eqnarray}
Inserting this into Eq.~(\ref{MPeqn}) yields
\begin{eqnarray}
&~&S_{L L^\prime R R^\prime I_s J_s}  = \sum_{\cal J} \,\left[
\begin{array}{ccc} 1 & R^\prime & I_s
\\ R & 1 & {\cal J} \end{array} \right] \left[ \begin{array}{ccc}
L^\prime & R^\prime & J_s \\ R & L & {\cal J} \end{array} \right]
s_{{\cal J} L L^\prime}^t +  O(N_c^{-1}) ,
\label{IequalsJ1}
\end{eqnarray}
where
\begin{eqnarray}
&{}& s_{{\cal J} L L^\prime}^t \equiv \frac{(-1)^{2{\cal J}} \,
[{\cal J}] }{3([L][L^\prime])^{1/2}} \sum_K \, [K] \left[
\begin{array}{ccc} 1 & L^\prime & K
\\ L & 1 & {\cal J} \end{array} \right] s_{K L L^\prime} ,
\label{IequalsJ2}
\end{eqnarray}
and for simplicity of presentation we have replaced the standard
$6j$ symbol with a new symbol denoted by square brackets that
folds in useful phase factors and overall constants, but retains all
the usual triangle rules:
\begin{equation}
\left\{\begin{array}{ccc} a & b & e \\ c & d & f
\end{array} \right\} \equiv
\frac{(-1)^{-(b+d+e+f)}}{([a][b][c][d])^{1/4}}
\left[\begin{array}{ccc} a & b & e \\ c & d & f \end{array}
\right].
\end{equation}
Use of the modified $6j$ symbols, henceforth called $[6j]$ symbols,
leaves Eqs.~(\ref{IequalsJ1})--(\ref{IequalsJ2}) much more compact
than the corresponding expressions using ordinary $6j$ symbols.

The energy-dependent function $s_{{\cal J} L L^\prime}^t$ is the
$t$-channel reduced amplitude; it depends only on the pion orbital
momentum and the SU(2) index ${\cal J}$.  Applying triangle rules
to each of the $[6j]$ symbols in Eq.~(\ref{IequalsJ1}) reveals the
physical significance of ${\cal J}$.  The first $[6j]$ symbol
implies that ${\cal J}$ is the total isospin ($I_t$) exchanged
between the meson and baryon in the $t$-channel, while the second
implies that ${\cal J}$ is the total angular momentum ($J_t$)
exchanged in the $t$-channel.  Together, they demand the equality
of isospin and angular momentum in the $t$-channel exchange, in
accordance with the $I_t = J_t$ rule of Mattis and
Mukerjee~\cite{MM,Mat3}.

While this rule was originally derived in the Skyrme model, it was
shown to be a result of large $N_c$ QCD by Kaplan and
Manohar~\cite{KapMan} through the model-independent spin-flavor
approach based on large $N_c$ consistency conditions (which in
turn follows from the pioneering work of
Refs.~\cite{GS,DM,DJM}).  They demonstrate that the matrix
element of a general $n$-quark operator $\hat{O}^{(n)}_{I_0,
J_0}$ with baryon number equal to zero, isospin $I_0$ and spin
$J_0$ scales as:
\begin{equation}
\langle B^\prime|\hat{O}^{(n)}_{I_0, J_0}/N^n_c|B \rangle \lesssim
1/N_c^{|I_0-J_0|}.\label{me}
\end{equation}
The significance of this result becomes manifest when one realizes
that the ``quarks'' in this derivation need not be associated with
dynamical quarks in any particular quark model.  Rather, they
merely reflect fields transforming according to the fundamental
representation of the contracted SU($2 N_f$) symmetry~\cite{DJM}.
Thus, the rule applies to all baryon matrix elements.  The
operator that connects the pions and baryons in a $t$-channel
exchange, from the point of view of the baryon, is simply a
single current insertion that couples to external pions; its
matrix element between baryon states then qualifies as the type
described above.  One sees from Eq.~(\ref{me}) that the largest
contribution to the scattering comes from matrix elements with
$I_t = J_t$; thus, the famed $I_t = J_t$ rule is a direct result
of large $N_c$ QCD without model input. Since
Eq.~(\ref{IequalsJ1}) is the most general form for a scattering
amplitude consistent with the $I_t = J_t$ rule, we have
established that Eq.~(\ref{MPeqn}) is a model-independent, large
$N_c$ QCD result. This general argument was originally presented
in Ref.~\cite{CL}.

The rederivation of Eq.~(\ref{MPeqn}) by nonsolitonic means is of
only modest interest.  However, the crucial point is that the
general large $N_c$ derivation can be extended to higher order in
$1/N_c$.  The method by which one extends the earlier LO results
to NLO is clear: Since the LO $t$-channel constraint on the
amplitudes, $|I_t-J_t|=0$, implies Eq.~(\ref{MPeqn}), the first
linearly independent $1/N_c$ correction arises from $t$-channel
amplitudes with $|I_t-J_t|=1$, since Eq.~(\ref{me}) implies that
such amplitudes are the ones suppressed by a single factor
$1/N_c$. All $t$-channel amplitudes with $|I_t-J_t|> 1$ can be
excluded at NLO since the suppression is $1/N_c^2$ or more.  As
we now show, an expansion to this order remains predictive since
only two $t$-channel amplitudes with $|I_t-J_t|=1$ appear.

Writing the pion-baryon partial-wave amplitude in terms of reduced
$t$-channel amplitudes and including the first subleading
contributions from amplitudes with $|I_t-J_t|=1$ generalizes
Eq.~(\ref{IequalsJ1}):
\begin{eqnarray}
&~& S_{LL^\prime R R^\prime I_s J_s} = \sum_{\cal J} \left[
\begin{array}{ccc} 1 & R^\prime & I_s\\ R & 1 & {\cal J}
\end{array} \right] \left[
\begin{array}{ccc} L^\prime & R^\prime & J_s \\ R & L & {\cal J} \end{array}
\right] s_{{\cal J} L L^\prime}^t \nonumber \\ &~&
-\frac{1}{N_c}\sum_x \left[ \begin{array}{ccc} 1 & R^\prime &
I_s\\ R & 1 & x
\end{array} \right] \left[
\begin{array}{ccc} L^\prime & R^\prime & J_s \\ R & L & {x+1} \end{array}
\right] s_{x L L^\prime}^{t(+)} - \frac{1}{N_c}\sum_y \left[
\begin{array}{ccc} 1 & R^\prime & I_s\\ R & 1 & y
\end{array} \right] \left[
\begin{array}{ccc} L^\prime & R^\prime & J_s \\ R & L & {y-1} \end{array}
\right] s_{y L L^\prime}^{t(-)} +  O(N_c^{-2}) , \label{MPplus}
\end{eqnarray}
where the $s_{x L L^\prime}^{t(\pm)}$ functions are the reduced
$t$-channel amplitudes corresponding to $s_{{\cal J} L L^\prime}^t$
for the two possible ways of combining $I_t$ and $J_t$ such that
$|I_t-J_t|=1$.  In Sec.~\ref{lin} this formula is used to derive
linear relations among partial-wave amplitudes for $\pi
N\rightarrow\pi N$ and $\pi N\rightarrow\pi\Delta$ at NLO.  As noted
previously, any relations that depend explicitly on the higher order
amplitudes $s_{x L L^\prime}^{t(\pm)}$ have essentially the same
predictive power as the LO relations.  However, if gold-plated
relations can be found in which the effects of the $s_{x L
L^\prime}^{t(\pm)}$ cancel, then we have predictions that hold at NLO
and thus are expected to describe nature far better than the generic
LO relations of Eqs.~(\ref{LR1})--(\ref{LR6}).

\section{Linear Relations \label{lin}}

Before deriving gold-plated NLO linear relations, it is helpful to
discuss restrictions on the reduced amplitudes and the pion
angular momentum due to symmetry.  Time-reversal invariance of
the scattering process dictates that the $S$ matrix is symmetric
under the exchange of initial and final states (characterized by
$L R$ and $L^\prime R^\prime$, respectively).  We see that the
symmetry properties of the $[6j]$ symbols (inherited from the
usual $6j$ symbols) imply that they are invariant under this
exchange.  Thus, all types of reduced amplitudes must also be
symmetric ({\it e.g.}, $s^t_{{\cal J} L L^\prime}=s^t_{{\cal J}
L^\prime L}$) in order to maintain the symmetries of QCD.  The
$[6j]$ symbols also encode important restrictions on $\Delta L
\equiv |L^\prime - L|$.  For $\pi N \rightarrow \pi N$, the
allowed change is $\Delta L=0,1$; while for $\pi N \rightarrow
\pi \Delta$, the allowed change is $\Delta L=0,1,2$.  The $\Delta
L=1$ possibility is eliminated by parity conservation since
$P=(-1)^{L+1}=(-1)^{L^\prime+1}$.  To summarize, the permitted
cases are $\Delta L=0$ for $\pi N \rightarrow \pi N$ and $\Delta
L=0, 2$ for $\pi N \rightarrow \pi \Delta$.

Let us first consider the reactions $\pi N \rightarrow \pi N$ and
$\pi N \rightarrow \pi \Delta$ when the pion orbital angular
momentum is unchanged: $L=L^\prime$.  There are eight physical
amplitudes corresponding to the different ways to add the spin and
isospin of the pion and the nucleon in the two reactions:
$I_s=\frac{1}{2}, \frac{3}{2}$ and $J_s=L \pm \frac{1}{2}$.  We
can expand these in terms of seven reduced amplitudes: three LO
and four first order in $1/N_c$.  Therefore, there is only one
relation among the physical amplitudes with all references to the
NLO reduced amplitudes eliminated.  This gold-plated relation is:
\begin{eqnarray}
S^{\pi N}_{L,1,-1}-S^{\pi N}_{L, 1,+1} & = & \sqrt{\frac{2L-1}{L+1}}
S^{\pi \Delta}_{L,1,-1}+\sqrt{\frac{2L+3}{L}}
S^{\pi \Delta}_{L,1,+1}\label{DN1} +\left[
\frac{2}{3}\left(S^{\pi N}_{L, 3,-1}-
S^{\pi N}_{L,3,+1}\right)+\frac{1}{3}\left(S^{\pi N}_{L,1,-1}-
S^{\pi N}_{L,1,+1}\right)\right. \nonumber \\ &&
\left.+\frac{1}{3}\sqrt{\frac{5}{2}}\left(\sqrt{\frac{2L-1}{L+1}}
S^{\pi \Delta}_{L,3,-1}+\sqrt{\frac{2L+3}{L}}
S^{\pi \Delta}_{L,3,+1}\right)-\frac{5}{6}\left(\sqrt{\frac{2L-1}{L+1}}
S^{\pi \Delta}_{L,1,-1}+\sqrt{\frac{2L+3}{L}}
S^{\pi \Delta}_{L,1,+1}\right)\right] \nonumber \\ && + O(N_c^{-2}).
\end{eqnarray}
The first four terms resemble one of the original MP relations,
Eq.~(\ref{LR6}), but there is a correction term in the square
brackets.  Note that this correction term itself vanishes as $N_c
\rightarrow \infty$ after substituting in
Eqs.~(\ref{LR1})--(\ref{LR4}). However, the $1/N_c$ corrections
to the terms in the square bracket from
Eqs.~(\ref{LR1})--(\ref{LR4}) precisely cancel the corrections to
Eq.~(\ref{LR6}), yielding a result that holds to $O(1/N_c^{2})$.
Equation~(\ref{LR6}) empirically works rather well, and we defer
a discussion of the possible effects of the correction term to
Sec.~{\ref{exp}}.

Now we consider the reactions for which the pion orbital angular
momentum is changed by two units, $L=L^\prime\pm2$; the symmetry
arguments given above restrict this case to the $\pi N
\rightarrow \pi \Delta$ reaction.  There are four physical
amplitudes for this case.  They can be expressed in terms of two
reduced amplitudes: one leading order and one first order in $1/N_c$.
This implies the existence of two gold-plated linear relations:
\begin{eqnarray}
\sqrt{L+1} \, S^{\pi \Delta}_{L,L+2,1,+1} & = & - \sqrt{L+2} \,
S^{\pi \Delta}_{L+2,L,1,-1} + O(N_c^{-2}), \label{DN2}\\
\sqrt{L+1}S^{\pi \Delta}_{L,L+2,3,+1} & = & - \sqrt{L+2} \,
S^{\pi \Delta}_{L+2,L,3,-1}  + O(N_c^{-2}). \label{DN3}
\end{eqnarray}
These resemble two of the MP relations [{\it cf.}\
Eqs.~(\ref{LR5a}), (\ref{LR5b})].  However, we have now shown
that they hold at NLO, and thus are gold- rather than
silver-plated. Thus, to the extent that the $1/N_c$ expansion
applies to these observables, one expects that these relations
hold far better than the generic silver-plated LO predictions. As
discussed in the following section, we show that this is, in
fact, true.

\section{Experimental Tests \label{exp}}

In principle, all three linear relations derived in Sec.~\ref{lin}
(for each allowed value of $L$) can be tested by comparison with
available experimental data.  The numbers used result from
partial-wave analysis applied to raw data from experiments in
which pions are scattered off nucleon targets.  An important
feature complicates our task: All of the gold-plated relations
involve the reaction $\pi N \rightarrow \pi \Delta$.  While the
extraction of partial-wave amplitudes for the $\pi N \rightarrow
\pi N$ reaction from the large amount of reliable data is
essentially straightforward, the extraction of partial-wave
amplitudes for $\pi N \rightarrow \pi \Delta$ is complicated by
the fact that the $\Delta$ decays strongly to $\pi N$.  The $\pi
N \rightarrow \pi \Delta$ partial waves must be extracted in the
context of a model that distinguishes events in the observed
reaction $\pi N \rightarrow \pi \pi N$ that pass through an
intermediate $\Delta$ resonance and which do not.  Therefore, the
$\pi N \rightarrow \pi \Delta$ partial-wave amplitude data
necessarily contains some model dependence, making it somewhat
less reliable.  Due to this uncertainty, much less attention has
been paid to these reactions, and the set of analyzed data is far
more sparse. Fortunately, the $\Delta$ is an extremely prominent
resonance (understandable in the context of large $N_c$), and
hence the model dependence should be rather modest.

For the comparison presented below we use results from the analysis of
Manley, Arndt, Goradia, and Teplitz~{\cite{Arndt}}, which is readily
available through the SAID program at GWU~{\cite{SAID}}.  The analysis
is presented in terms of the $T$ matrix [${\bf T}\equiv({\bf S}-{\bf
1})/2i$] rather than the $S$ matrix.  This causes no complications,
since any extra factors and terms cancel in our formulas.  The results
of Ref.~\cite{Arndt} are presented in terms of the center-of-mass
energy $W$ of the $\pi N$ system.

We first consider Eq.~(\ref{DN1}) and restrict attention to $1
\leq L \leq 3$.  The lower bound is an elementary consequence of
angular momentum conservation, while the upper bound reflects
limitations of the available data.  Even with this restriction we
see that for each $L$, Eq.~(\ref{DN1}) requires partial-wave
amplitudes that are, unfortunately, not available in the data
set.  For example, the amplitudes $PP_{31}$, $PP_{13}$,
$DD_{33}$, and $FF_{17}$ (the notation is $L L^{\prime}_{2I 2J}$)
are not given.  MP in their LO comparisons were able to
circumvent this problem by rewriting the unknown amplitudes in
terms of known ones using formulas~Eqs.~(\ref{LR3}),
(\ref{LR4}).  We have no such luxury; inserting~Eqs.~(\ref{LR3}),
(\ref{LR4}) into our gold-plated relations simply converts them
to silver-plated relations.  We make no assumptions about these
unknown amplitudes and thus cannot test the validity of
Eq.~(\ref{DN1}) at the present time.

We now consider Eqs.~(\ref{DN2}), (\ref{DN3}).  Fortunately,
there is sufficient analyzed data to study these relations,
provided one restricts attention to the $L=0$ case.  It is
instructive to contrast the quality of the agreement of these
gold-plated NLO relations with the $L=0$ silver-plated LO
relations Eqs.~(\ref{LR5primea}), (\ref{LR5primeb}), since both
sets involve only the $\pi N \rightarrow \pi \Delta$ amplitudes.
We view the loss of predictive power due to the need to identify
the $\Delta$ in the final state as a comparable systematic
uncertainty for the two classes of relations.  Our predictions
are as follows:
\begin{eqnarray}
SD_{11}&=&-\sqrt{2}\, DS_{13} + O(1/N_c^2) , \label{testDN1}\\
SD_{31}&=&-\sqrt{2}\, DS_{33} + O(1/N_c^2) , \label{testDN2}\\
SD_{11}&=&+\sqrt{20}\, DS_{33} + O(1/N_c) , \label{testMP1}\\
SD_{31}&=&+\frac{1}{\sqrt{5}}\, DS_{13}+ O(1/N_c)
\label{testMP2},
\end{eqnarray}
where the first two relations are the gold-plated NLO relations
(Fig.~\ref{fig1}) and the second two are the silver-plated LO
relations (Fig.~\ref{fig2}).

\begin{figure}
\includegraphics[0in,0in][6.5in,3in]{fig1.eps}
\caption{Experimentally determined $\pi N \rightarrow \pi \Delta$
amplitudes $SD_{11}$ and $SD_{31}$ compared to the predictions of
Eqs.~(\ref{testDN1}), (\ref{testDN2}).  In plots (a) and (b), the
closed circle ($\bullet$) is $SD_{11}$ and the box ($\Box$) is
$-\sqrt{2}\, DS_{13}$.  In plots (c) and (d), the open circle
($\circ$) is $SD_{31}$ and the diamond ($\diamond$) is
$-\sqrt{2}\, DS_{33}$.  The data is provided by
SAID~\cite{SAID}.} \label{fig1}
\end{figure}

\begin{figure}
\includegraphics[0in,0in][6.5in,3in]{fig2.eps}
\caption{Experimentally determined $\pi N \rightarrow \pi \Delta$
amplitudes $SD_{11}$ and $SD_{31}$ compared to the predictions of
Eqs.~(\ref{testMP1}), (\ref{testMP2}).  In plots (a) and (b), the
closed circle ($\bullet$) is $SD_{11}$ and the diamond ($\diamond$) is
$+\sqrt{20}\, DS_{33}$.  In plots (c) and (d), the open circle
($\circ$) is $SD_{31}$ and the box ($\Box$) is $+1/\sqrt{5}\,
DS_{13}$.  The data is provided by SAID~\cite{SAID}.} \label{fig2}
\end{figure}

It is immediately apparent that the gold-plated relations agree with
experiment considerably better than their silver-plated analogs.  For
the gold-plated relations the gross structure of the amplitudes is
clearly discerned on both the left- and right-hand sides of the
relation.  In contrast, the silver-plated relations are much less
robust in describing the data.

\section{Conclusion}

We have demonstrated the utility and power of the large $N_c$
expansion for describing pion-nucleon scattering.  It has made a
number of nontrivial predictions that can be tested with
experimental data.  The expansion in powers of $1/N_c$ allows one
to compare predictions holding at different orders, and the
quality of the agreement for the $O(1/N_c^2)$ relations is
markedly better than the $O(1/N_c)$ relations.  It is unfortunate
that sufficient analyzed data does not exist for our gold-plated
relation Eq.~(\ref{DN1}).  In principle, the relevant $\pi N
\rightarrow \pi \Delta$ partial-waves might be extracted from the
raw data.  However, this requires a formidable (and
model-dependent) analysis.  Previously there was, perhaps, little
motivation to carry out this analysis, but in light of these large
$N_c$ predictions the incentive is now more compelling.

It is exciting to see that some rather complicated features of QCD,
such as the $\pi N \rightarrow \pi \Delta$ reaction, can be understood
semi-quantitatively in terms of rather simple microscopic
considerations based on large $N_c$.

In principle, our method can be applied again to derive the
$1/N_c^2$ terms in the $S$ matrix expansion.  However, we note
that such a procedure is of minimal utility for describing
pion-nucleon scattering in the physical $N_c=3$ world.  The
resulting triangle rules appearing in the $1/N_c^2$ corrections,
applied to terms with a nucleon ($R=\frac 1 2$), cannot be
satisfied for any baryon in the $R^\prime = I=J$ multiplet of the
large $N_c$ world; this forces the $6j$ symbols to vanish, thus
terminating the expansion.  Therefore, it appears that we have
exhausted the number of experimentally accessible gold-plated
relations in pion-nucleon scattering, and we see that there are no
``super''-gold-plated relations that hold at
next-to-next-to-leading order\@.

This approach can clearly be extended to other processes.  For
example, one may relate partial waves in Compton scattering, electron
scattering, and pion-electron production, or photoproduction.  We
defer such considerations to later work.

\section*{Acknowledgments}

D.C.D.\ would like to thank Richard Arndt for his assistance with
collecting and interpreting the data from SAID\@. D.C.D.\ would
also like to thank E.A.~Rogers for assistance in making the
plots.  The work of T.D.C., D.C.D., and A.N.\ was supported in
part by the U.S.\ Department of Energy under Grant No.\
DE-FG02-93ER-40762.  The work of R.F.L.\ was supported by the
National Science Foundation under Grant No.\ PHY-0140362.

\end{document}